\RequirePackage{fix-cm}
\documentclass[twocolumn,epjc3]{svjour3}  
\smartqed  % flush right qed marks, e.g. at the end of the proof
\RequirePackage{graphicx}
\usepackage{hyperref}
\hypersetup{colorlinks,citecolor=blue}

\usepackage{color}
\usepackage{amsmath}
\usepackage{amssymb}
\journalname{Eur. Phys. J. C}
\begin{document}

\title{Testing of $\kappa(\mathcal{R},\mathcal{T})-$gravity through gravastar configurations}


\author{G  R  P  Teruel\thanksref{e1,addr1}
        \and
        Ksh. Newton Singh \thanksref{e5,addr2} 
        \and
        Tanmoy Chowdhury \thanksref{e4,addr3}
         \and
        Farook Rahaman \thanksref{e2,addr3} 
        \and
        Monimala Mondal \thanksref{e3,addr3} 
}

\thankstext{e1}{e-mail: gines.landau@gmail.com}
\thankstext{e5}{e-mail: ntnphy@gmil.com}
\thankstext{e4}{e-mail: tanmoych.ju@gmail.com}
\thankstext{e2}{e-mail: rahaman@associates.iucaa.in}
\thankstext{e3}{e-mail: monimala.mondal88@gmail.com}


\institute{Departamento de Matematicas, IES Carrus, Elche 03205, Alicante, Spain. \label{addr1}
\and
Department of Physics, National Defence Academy, Khadakwasla, Pune-411023, India. \label{addr2}
\and
Department of Mathematics, Jadavpur University, Kolkata-700032, West Bengal, India \label{addr3} }

\date{Received: date / Accepted: date}

\maketitle

\begin{abstract}
In  this  article,  we are reporting for the first time the existence of gravastar configurations in the framework of  $\kappa(\mathcal{R},\mathcal{T})-$ gravity, which can be treated as an  alternative to a black hole (Mazur and Mottola). This strengthens how much this new gravity theory may be physically demanding to the gravity community in the near future. We first develop the gravastar field equations for a generic $\kappa(\mathcal{R},\mathcal{T})$ functional and then we study four different models within this theory. We find that the solutions for the interior region are regular everywhere regardless of the exact form of the $\kappa(\mathcal{R},\mathcal{T})$ functional. The solutions for the shell region indicate that two of the four models subjected to the study are physically feasible. In addition, the junction conditions are considered at each interface by using the Lanczos equations that yield the surface density and pressure at the thin shell. We investigate various characteristics of the gravastar structure such as the proper length, energy, and entropy of the spherical distribution.

\end{abstract}

\section{Introduction}
The hierarchy of principles in physics is an old debate that should be revived from time to time to ask ourselves if we are going in the right direction. On the one hand, there are foundational physical principles, such as the principle of inertia in classical mechanics, the uncertainty principle in quantum mechanics, or the principle of relativity that directly founded or contributed to establishing new scientific theories. On the other hand, there are the auxiliary principles, which are not as fundamental as the first ones, but help to strengthen our theories once they have been established by the foundational principles. It is natural to wonder: To which of these two categories does the principle of least action belong? At first glance, one would answer that the least action principle should be an undisputed part of the first group, i.e., a foundational one due to its ubiquitous presence in our current theories. However, let's carefully examine the history of physics. We will appreciate that the variational principles, despite their enormous success and importance, were not founding principles, in the sense that they did not establish new scientific paradigms, although they turned out to be versatile and flexible enough to know how to adapt to every paradigm shift in theoretical physics.

In particular, if we examine the birth of the two major revolutions of the twentieth century, the theories of relativity and quantum mechanics, we will see that the least action principle had a secondary role \cite{reig,Hei}. In fact, other principles were the foundational building blocks of these theories \cite{Ein,Hei2}. The action principle was adapted later to them and arrived to enrich their formalism. Analogously, neither of the two great classical field theories was discovered by means of the least action principle. Indeed, Classical Electrodynamics (CE) was fulfilled with the inclusion of a source term directly into the field equations (the Maxwell displacement current). General relativity (GR) was also first derived in this way, after adding the key trace term to the field equations \cite{hil15,ren07}. In both cases, the least action principle came to strengthen the formalism after other more fundamental principles had established the theories.

These historical lessons persuade us that the search for possible Non-Lagrangian theories, that is, theories not based (at least initially) on the variational method, should not be ruled out a priori.

In 2001, Mazur and Mottola \cite{Mazur01,Mazur04}  proposed a very interesting and encouraging stellar model, named gravitational vacuum star (or gravastar) as an alternating black hole structure. In recent days, the study on gravastar has become a considerably interesting topic as an alternating black hole solution. Mazur and Mottola \cite{Mazur01,Mazur04} first introduced a novel solution for the extreme point of gravitational collapse of cold and dark matter, a neutral system, and compact stars. Therefore, this is an extended form of Bose-Einstein condensation whose radius is similar to the Schwarzschild radius. The total design of a gravastar can be defined by means of 3 different zones with distinct EoS:\\

I. Interior: $ 0<r<r_1 $  ~~( $ p=-\rho $),\\

II. Shell: $ r_1<r<r_2 $  ~~($p=\rho$),\\

III. Exterior: $r_2<r$   ~~($p=\rho=0$).\\

Several researchers have already studied gravastar from different perspectives and scooped out a new field as an alternative to the black hole paradigm.
\cite{Mazur01,Mazur04,visser04,cattoen05,Bilic06,Lobo07,Rocha08,Horvat07,Nandi09,Usmani11,FR15}. Lack of observational evidence in Einstein's GR on the oscillating model of the universe alongside the presence of dark matter has forced a conceptual challenge to this theory \cite{Chan,C.F.C,R.CHAN,D.J}.

The repulsive nature of dark energy in GR requires considering an exotic fluid with a negative pressure to explain the cosmic speed-up. Many modified Lagrangian theories of gravity have been proposed to account for the accelerating phase of the universe. These theories are based on the generalization of the Einstein-Hilbert action {\it viz} $f(\mathbb{T},\mathcal{T})$ \cite{TH14}, $f(\mathbb{T})$ \cite{v.c}, $f(\mathcal{R},\mathcal{T})$ \cite{Harko}, EGB \cite{Lovelock}, $f(G)$ \cite{N.M.G}, $f(G,\mathcal{T})$ \cite{MS}, $f(Q, \mathcal{T})$ \cite{Y.X} gravity theories, etc, which earned wide attention in recent years.

A major role of modified gravity theories is to test the validity of the gravastar model by examining various realistic properties of these compact objects. Barzegar et al. \cite{H.B} explored the 3D AdS gravitational vacuum stars in the framework of gravity’s rainbow theory. Ghosh et al. \cite{S.G20} investigated gravastars in $f(\mathbb{T},\mathcal{T})$ gravity and enlarged its research in Rastall gravity as well \cite{S.G21}. Das et al. \cite{A.DAS17}, investigated the formation of gravastars in $f(\mathcal{R},\mathcal{T})$ gravity and inspect its theoretical efficiency. Enlarge the analysis Yousaf et al., recently explored gravastars in $f(\mathcal{R}, \mathcal{T}, R_{\mu\nu} T^{\mu\nu})$ gravity \cite{Z.Y20}. Bhatti et al. \cite{M.Z20} reviewed charged gravastar for spherically and cylindrically symmetric spacetimes and in $f(\mathcal{R}, G)$ gravity \cite{M.Z.B2020}. Extending it, Yousaf \cite{ZY20} constructed charged cylindrical gravastar-like structures in $f(\mathcal{R},\mathcal{T})$ gravity. Lobo and Garattini \cite{F.S} analyzed gravastar solutions in the frame of non-invariant geometry with their physical properties and characteristics. Bhatti et al. \cite{MZ} investigated the stability of this thin shell together with the thermodynamical stability of  locally isotropic gravastars with cylindrical space-time. Bhatti et al. \cite{MZ21} talked over the symmetric gravastar model within the framework of the modification of Gauss-Bonnet gravity. Das et al. \cite{ad20} discussed the  physically interesting and valid features within the framework of the $f(\mathbb{T})$ theory of gravity.

Horvet et al. \cite{Horvat} used dominant force conditions to initialize the stability of the gravastar after combining a sufficient external vacuum solution with an internal geometry and meeting some possible constraints. Also, they examined the formation of this type of compact object under the influence of electromagnetism. Rahaman and others \cite{rahaman} analyzed the electrically charged gravastar model considering the $(2 +1)-$ dimensional geometry. For a stable model of a spherically symmetric gravastar, they also explored some viable properties as well as energy components, length, and entropy of the gravastar. Ghosh et al. \cite{ghosh} investigated some new diagrams in the formation of these matters with and without electric charge in high-order  manifolds. Ghosh et al. \cite{Ghosh19} discussed gravastar layout in the framework of the Kuchowicz metric potential.  Ghosh et al. \cite{Sg 19} figure out the gravastar model under Einstein’s GR in ($3+1$) dimensions by incorporating the Karmarkar condition. Sumita et al. \cite{sumita} studied gravastars under Finslerian spacetime geometry. Sengupta et al. \cite{R.S20} investigated a gravastar configuration in RS Brane gravity, while Ray et al. \cite{S.RAY20} reviewed a very interesting paper on generalizations of gravastars in GR, including studies of higher and lower dimensional GR with physical insights like the Randall-Sundrum theory, and several modified gravity models such as $f(\mathcal{R}, \mathcal{T})$ theory or Rastall-Rainbow gravity. 
 
Recently, a modified theory was invented \cite{gr18} as $\kappa(\mathcal{R},\mathcal{T})-$ gravity. The conceptual framework of this theory is not based on the standard modified gravity program. Rather, its insight is inspired by Maxwell’s and Einstein’s ideas of including new possible potential terms in the field equations. In this sense, we must recall Maxwell's addition to Ampere's law of the displacement current term to complete the Electromagnetic field equations, and the introduction of the key term, $\frac{1}{2} \mathcal{R}\,g^{\mu\nu}$ by Einstein to consummate the GR field equations \cite{hil15,ren07}. Despite the fact that the variational principle is often regarded as our main tool to formulate a new physical theory (and its generalizations), ought not to have been placed at the same fundamental level as other truly first principles, like the equivalence principle and the principle of general covariance, that are the two core principles of GR. In spite of this, the variational method has reached the pinnacle in our hierarchy of physical principles, becoming dominant to the extent of being the starting point for any modification of GR.
Indeed, the vast majority of the modified gravity theories that are available in the market, like the examples mentioned above, are Lagrangian in the sense that they arise from the generalization of the Einstein–Hilbert action. The overpopulation of alternatives suggests that some other fundamental principle (beyond the equivalence principle and the principle of general covariance) is required to guide us trough the deep jungle of possible Lagrangian theories. The fact is, we have no reason to assure that general symmetries and general conservation laws can always exist (in their current form) in the final theories of nature, which moves us to think that the Non-Lagrangian proposals should also be explored. In fact, according to Hojman \cite{Hoj}, the use of Lagrangian and Hamiltonian functions is not an essential requirement to derive conservation theorems.

$\kappa(\mathcal{R},\mathcal{T})-$gravity is a recent proposal, and so its implications have not been explored well enough. However, for the past few years, some works are dedicated to  studying its cosmic aspects. In particular,  Pradhan and Ahmed \cite{ahm22} showed that the theory can deal with the current scenario of an accelerating universe, claiming that the theory requires a very small value of the cosmological constant, which agrees with observations. Pradhan et al. \cite{arc22} further extended this study to include a more complete cosmological scenario, while Dixit et al. \cite{arc23} studied the thermodynamics of the expansion of the cosmos in the context of this theory. Sarkar et al. \cite{sark19}, explored a model of wormhole in $\kappa(\mathcal{R},\mathcal{T})-$ gravity, while Teruel et al. \cite{GRP} introduced the first internal solutions representing compact stars in $\kappa(\mathcal{R},\mathcal{T})$-gravity by solving the field equations in isotropic coordinates. Also very recently, Taṣer and Dogru \cite{Tas23} investigated whether the Krori-Barua model produces valid results in this theory. All these papers have one remarkable common feature: they were devoted to studying only a particular case of the theory with $\kappa(\mathcal{R},\mathcal{T}) \equiv \kappa(\mathcal{T})=8\pi-\lambda\mathcal{T}$.

 Up to now, nobody studied gravastar solutions in $\kappa(\mathcal{R},\mathcal{T})-$ gravity, it would therefore be interesting to investigate and discuss whether this theory can support a consistent gravastar configuration from a physical point of view. In this sense, four different models, i.e., four different choices of the running $\kappa(\mathcal{R},\mathcal{T})-$ gravitational constant are analyzed in this investigation. Therefore, we don't restrict ourselves only to the $\kappa (\mathcal{T})$ choice like all the previous works, but extend the study to the more general possible case. Indeed, once derived the equations for a generic $\kappa(\mathcal{R},\mathcal{T})$ functional, we see how the theory would look for specific cases.

The structure of the paper is as follows. We start in section \ref{sec2} with the basic framework of the theory. Section \ref{sec3} deals with the field equations derived from the Morris-Thorne metric for the general $\kappa(\mathcal{R},\mathcal{T})$ case. The solutions for the interior region of the gravastar are discussed in section \ref{sec4}. The shell region of the gravastar is analyzed in section \ref{sec5}, where we study the physical acceptability of 4 different models within the theory. Entropy within the shell is addressed in section \ref{sec6}. The junction interface and surface stresses are the subjects of section \ref{sec7}. The brief section \ref{sec8} deals with the surface redshift within the thin shell. Finally, in section \ref{sec9}, we summarize the findings and discuss the conclusions of the work.

\section{Framework of $\kappa(\mathcal{R},\mathcal{T})-$ gravity }\label{sec2}
The structure of the theory depends on the following field equations
\begin{equation}\label{FieldEquations}
R_{\mu\nu}-\frac{1}{2}\mathcal{R}~g_{\mu\nu}-\Lambda g_{\mu\nu}=\kappa(\mathcal{R},\mathcal{T})\,T_{\mu\nu},
\end{equation}
where $R_{\mu\nu}$ is the Ricci tensor, $g_{\mu\nu}$ is the space-time metric, $\Lambda$ corresponds to the cosmological constant, $T_{\mu\nu}$ is the stress-energy tensor of the material content, while $\kappa(\mathcal{R},\mathcal{T})$ is a generalization of Einstein's gravitational constant, that we extend to the status of a function of the traces $\mathcal{T} \equiv g_{\mu\nu}T^{\mu\nu}$, and $\mathcal{R} \equiv g_{\mu\nu}R^{\mu\nu}$. The introduction of the functional  $\kappa(\mathcal{R},\mathcal{T})$, means that we inspect the possibility of a running gravitational constant, although not at the level of the variational method. The possibility of considering a variable Einstein's gravitational constant in the action was studied many years ago by Brans and Dicke \cite{brans1,brans2}, and the resulting theory is quite different from (\ref{FieldEquations}).

The field equations (\ref{FieldEquations}) imply the non-covariant conservation of $T_{\mu\nu}$. Indeed, since the divergence of the left-hand side of Einstein's field equation vanishes, we have
\begin{equation}\label{nonconservation}
\nabla{^\nu}\Big[\kappa(\mathcal{R},\mathcal{T})\,T_{\mu\nu}\Big]=0.
\end{equation}
Then, this non-conservation of the $T_{\mu\nu}$ can be expressed, for $\kappa(\mathcal{R},\mathcal{T})\neq 0$, as
\begin{equation}\label{nonconservation2}
\nabla^{\nu}T_{\mu\nu}=-\frac{\nabla^{\nu} \kappa(\mathcal{R},\mathcal{T})}{\kappa(\mathcal{R},\mathcal{T})}~T_{\mu\nu}, ~~~~~\forall~~~~\kappa(\mathcal{R},\mathcal{T})\neq0. 
\end{equation} 
When $\kappa(\mathcal{R},\mathcal{T})=0$, the right-hand side of Einstein's field equations vanishes. This interesting case probably may happen in some specific models in the early universe for sufficiently high densities, and it will imply an exponential expansion driven by a cosmological constant. Of course, the theory can be recast into a conservative form by defining a new (effective) stress-energy tensor
\begin{equation}
S_{\mu\nu}=\kappa(\mathcal{R},\mathcal{T})\, T_{\mu\nu}.
\end{equation}
The field equations then take the form
\begin{equation}
R_{\mu\nu}-\frac{1}{2}\mathcal{R}~g_{\mu\nu}-\Lambda\, g_{\mu\nu}=S_{\mu\nu},
\end{equation}
and the Bianchi identities imply that Eq. (\ref{nonconservation}) is
\begin{equation}
\nabla^{\nu}S_{\mu\nu}=0.
\end{equation}
Then, this theory has the same formal structure as GR, with a non-trivial modification of the material content. Examples of famous non-conservative gravitational theories are Rastall's gravity \cite{Ras}, or the Lagrangian $f(\mathcal{R},\mathcal{T})$ theory by Harko et al. \cite{Har}.

Lindblom and Hisock \cite{Lind}, and also Visser \cite{Viss} more recently, criticized some non-conservative gravity theories (in particular the Rastall case), claiming that they are formally identical to Einstein's gravity and that one can always build a conserved effective stress-energy tensor $T^{eff}_{\mu\nu}$, which is constructed solely from the matter sources. Furthermore, they underline the fact that, if the effective stress-energy tensor of the matter sources does not depend on the space-time curvature, (such is the case of Rastall gravity), then these theories will not represent truly alternative theories of gravity, reducing the question of the non-conservation of the stress-energy tensor to the domain of special relativity. Nevertheless, since the effective stress-energy tensor of $\kappa(\mathcal{R},\mathcal{T})$ gravity is not in general determined by the matter sources solely (it can have a dependence on the space-time curvature as well via the trace $\mathcal{R}$), we conclude that $\kappa(\mathcal{R},\mathcal{T})$ theory represents a true generalization of GR and not only a redefinition of its matter sector.
Some relevant features of the theory are the following:
\begin{itemize}
\item The modification procedure involves the right-hand side of Einstein's field equations, that is, only the material content sector is generalized. Therefore, the equations will be second order in the metric coefficients, and the theory will be free of typical instabilities that plagued many of the higher-order gravitational theories.
\item Since the pure geometrical sector is the same as GR, the theory boils down to GR in the absence of matter sources.
\item The dependence on $\mathcal {T}$ means that, for particular cases of the type $\kappa(\mathcal{T})=8\pi-\lambda\mathcal{T}$, or for more general functional\footnote{This is also true for a generic functional $\kappa(\mathcal{R},\mathcal{T})=8\pi-\gamma \mathcal{R}f(\mathcal{T})$, with $\gamma$ a real constant and $f(\mathcal{T})$ an arbitrary polynomial function of $\mathcal{T}$.}, that directly couples the $\mathcal{R}$ and $\mathcal{T}$ traces, the theory would predict the same physics as GR when coupled to standard (traceless) electromagnetic fields. In such cases, only significant departures are expected for non-linear electrodynamics, where $\mathcal {T}\neq0$.
\end{itemize}

\section{Field Equations in  $\kappa(\mathcal{R},\mathcal{T})-$ gravity}\label{sec3}
The interior of the gravastar is taken to be as Morris-Thorne Spacetime given by 
\begin{small}
\begin{eqnarray}
ds^{2}=-e^{2f}dt^{2}+\left[1-\frac{b}{r}\right]^{-1}dr^{2} +r^{2}(d\theta^{2}+\sin^{2}\theta \,d\phi^{2})   . 
\end{eqnarray} 
\end{small}
For $\Lambda=0$, the Einstein field equations for a general $\kappa(\mathcal{R},\mathcal{T})$ model will be
 \begin{eqnarray}
\rho(r)~\kappa(\mathcal{R},\mathcal{T}) &=& \frac{b'(r)}{r^2},  \\
p(r)~\kappa(\mathcal{R},\mathcal{T}) &=& \frac{2f'(r)}{r}\left(1-\frac{b(r)}{r}\right)-\frac{b(r)}{r^3}.
\end{eqnarray}
On the other hand, the non-covariant conservation of the stress-energy tensor given by Eq. \eqref{nonconservation} acquires the form
\begin{small}
\begin{equation}\label{nonconservation3}
p(r)~\frac{d \kappa(\mathcal{R},\mathcal{T})}{dr}+\kappa(\mathcal{R},\mathcal{T})\left(\frac{d f(r)}{dr}\{\rho(r)+p(r)\}+\frac{d p(r)}{dr}\right)=0.
\end{equation}
\end{small}

\section{Regular Interior Solutions in   {$\kappa(\mathcal{R},\mathcal{T})-$} gravity with \NoCaseChange{$p=-\rho$}}\label{sec4}
In this section, we show that the interior solutions for a gravastar configuration are regular everywhere in $\kappa(\mathcal{R},\mathcal{T}) $ theory, regardless of the specific choice of the running gravitational constant. To see how this interesting property of the theory arises, notice that for $p(r)=-\rho(r)$, Eq. \eqref{nonconservation3} reduces to
\begin{equation}
\rho(r)~\frac{d \kappa(\mathcal{R},\mathcal{T})}{dr}+\kappa(\mathcal{R},\mathcal{T})~\frac{d \rho(r)}{dr}=0,
\end{equation}
which further takes the form
\begin{equation}
\frac{d}{dr}\Big[\kappa(\mathcal{R},\mathcal{T})~\rho(r)\Big]=0.
\end{equation}
This implies the relation
\begin{equation}\label{constant}
\kappa(\mathcal{R},\mathcal{T})\, \rho(r)=C,
\end{equation}
where $C$, the constant of integration. Therefore,  field equations reduce to
\begin{eqnarray}
C &=& \frac{b'(r)}{r^2},  \\
-C &=& \frac{2f'(r)}{r}\left(1-\frac{b(r)}{r}\right)-\frac{b(r)}{r^3},
\end{eqnarray}
from which we can obtain $b(r)$ and $f(r)$ as
\begin{eqnarray}
b(r) &=&\frac{C}{3}r^{3}+D,\\
f(r) &=& \frac{1}{2}\ln\Big(1-\frac{C}{3}r^{2}-\frac{D}{r}\Big).
\end{eqnarray}
Regularity at $r=0$ implies that we set $D=0$. At $r=0$, the core of the gravastar, we have that $f(r)=0$ and the metric functions become $g_{tt}=-1$, and $g_{rr}=1$. The density should also be constant and finite at the gravastar's interior. At first we observed that the curvature scalar is constant and non-divergent everywhere at the interior region. After a cumbersome calculation, we have arrived at
\begin{eqnarray}
&& \mathcal{R}=-2\Big(1-\frac{b(r)}{r}\Big)\Big[f^{\prime\prime}(r)+\frac{2f^{\prime}(r)}{r}+f^{\prime}(r)^{2}\Big]+\frac{2b^{\prime}(r)}{r^{2}} \nonumber \\
&& \hspace{1cm} +\frac{d}{dr}\left(\frac{b(r)}{r}\right)f^{\prime}(r)=4C.
\end{eqnarray}
Therefore, an important conclusion of our results is that the interior solution for the gravastar will be regular and well-behaved everywhere, regardless of the specific functional forms of 
$\kappa(\mathcal{R},\mathcal{T})$.

\subsection{Model $\kappa(\mathcal{R},\mathcal{T})=8\pi+\beta \mathcal{R}-\alpha \mathcal{T}$}
Let us consider the choice, $\kappa(\mathcal{R},\mathcal{T})=8\pi+\beta \mathcal{R}-\alpha \mathcal{T}$, where the pre-multiplier are the free parameters. For this particular case we have, using that $\rho(r)=-p(r)$ and the results above, that Eq. \eqref{constant} has the form
\begin{equation}\label{algebraic}
\Big[8\pi+4\beta C-4\alpha\rho(r)\Big]\rho(r)=C.
\end{equation}
This algebraic equation implies, as we mentioned before, that the density (and pressure) for the interior layer of the gravastar is uniform, and for $\alpha \neq 0$, it has the following possible values
\begin{eqnarray}
\rho &=& \frac{1}{2\alpha}\Big(\omega\pm\sqrt{\omega^{2}-\alpha C} \Big)=\rho_{1}\\
p &=& -\frac{1}{2\alpha}\Big(\omega\pm\sqrt{\omega^{2}-\alpha C} \Big)= p_{1},
\end{eqnarray}
where $ \omega $ is a constant given by $\omega=2\pi+\beta C $ and $\omega\geq\sqrt{\alpha C}$.

The gravitational mass at the interior can is found from
\begin{align}
 M(C_{1})= \int_0^{R_{1}=C_{1}} 4 \pi r^2\rho ~dr = \frac{4}{3}\pi C_{1}^{3}\rho_{1}.
\end{align}\label{mactive}

\subsection{Model $\kappa(\mathcal{R},\mathcal{T}) \equiv \kappa(\mathcal{T})=8\pi-\alpha \mathcal{T}$} 
The density and pressure for this $\kappa(\mathcal{R},\mathcal{T})$ choice can be computed by setting $\beta=0$ in the previous case. Taking into account that for this choice, the constant $\omega=2\pi+\beta C$ reduces to $\omega=2\pi$, we obtain
\begin{eqnarray}
\rho &=& \frac{1}{2\alpha}\Big(2\pi\pm\sqrt{4\pi^{2}-\alpha C} \Big)=\rho_{2}\\
p &=& -\frac{1}{2\alpha}\Big(2\pi\pm\sqrt{4\pi^{2}-\alpha C} \Big)=p_{2}.
\end{eqnarray}
Thus, the active gravitational follows from the direct computation
\begin{align}
 M(C_{1})= \int_0^{R_{1}=C_{1}} 4 \pi r^2\rho \, dr = \frac{4}{3}\pi C_{1}^{3}\rho_{2}.
\end{align}\label{mactive2}

\subsection{Model $\kappa(\mathcal{R},\mathcal{T}) \equiv \kappa(\mathcal{R})=8\pi+\beta \mathcal{R}$} 
The density and pressure for this specific choice can be computed by setting $\alpha=0$ in the case A. Then, Eq. \eqref{algebraic} reduces to 
\begin{equation}
\Big(8\pi+4\beta C\Big)\rho=C.
\end{equation}
Therefore, 
\begin{eqnarray}
\rho &=& \frac{1}{4(\frac{2\pi}{C}+\beta)}=\rho_{3}\\
p &=& -\frac{1}{4(\frac{2\pi}{C}+\beta)}=p_{3}.
\end{eqnarray}
The mass is obtained in the same fashion as models \eqref{mactive} and \eqref{mactive2}, i.e, by direct integration of the density in the range $r=0$ and $r=R_{1}$
\begin{align}
 M(C_{1})=   \int_0^{R_{1}=C_{1}} 4 \pi r^2\rho \,dr = \frac{4}{3}\pi C_{1}^{3}\rho_{3}.
\end{align}

\subsection{Model $\kappa(\mathcal{R},\mathcal{T})=8\pi-\gamma \mathcal{R}\mathcal{T}$}
For this direct coupling among matter and curvature terms, we have that the density, pressure and active gravitational mass are given by
\begin{eqnarray}
\rho &=& \frac{\pi}{4C\gamma}\Big(1\pm\sqrt{1-\frac{C^{2}\gamma}{{\pi}^2}}\Big) =\rho_{4}\\
p &=& -\frac{\pi}{4C\gamma}\Big(1\pm\sqrt{1-\frac{C^{2}\gamma}{{\pi}^2}}\Big)=p_{4},
\end{eqnarray}
\begin{align}
 M(C_{1})=   \int_0^{R_{1}=C_{1}} 4 \pi r^2\rho \,dr = \frac{4}{3}\pi C_{1}^{3}\rho_{4}.
\end{align}
where, $\gamma$ is a parameter of the model that should satisfy the constraint $\gamma \leq(\pi/C)^{2}$ in order to get positive density in the core.

\section{Shell Region \NoCaseChange{$p=\rho$}}\label{sec5}
This region of the gravastar supposedly contains a stiff fluid that obeys EoS $p=\rho$ or $v=\sqrt{dp/d\rho}=1$, which is the most stiff EoS known as {\it Zeldovich fluid}. This condition implies that the non-covariant conservation of the stress-energy tensor given by Eq. \eqref{nonconservation3} acquires the form
\begin{equation}\label{nonconservation4}
\rho(r)\,\kappa^{\prime}(\mathcal{R},\mathcal{T})+\kappa(\mathcal{R},\mathcal{T})\Big[2f^{\prime}(r)\rho(r)+\rho^{\prime}(r)\Big]=0.
\end{equation}
On the other hand, Einstein's field equations become
\begin{eqnarray}\label{EFE1}
\rho(r)\,\kappa(\mathcal{R},\mathcal{T}) &=& \frac{b^{\prime}(r)}{r^{2}}, \\
\rho(r)\,\kappa(\mathcal{R},\mathcal{T}) &=& \frac{2f'(r)}{r}\left(1-\frac{b(r)}{r}\right)-\frac{b(r)}{r^{3}}.
\end{eqnarray}
Since, in general, the running gravitational constant $\kappa(\mathcal{R},\mathcal{T})$ depends on the scalars $\mathcal{R}$, $\mathcal{T}$, which in turn are functions of $b(r)$, $f(r)$, $\rho(r)$.  To solve the unknown functions $b(r)$, $f(r)$ and $\rho(r)$ we need three equations, two from field equations and one conservation equation. From the last two equations we obtain
\begin{equation}\label{combined}
\frac{b^{\prime}(r)}{r^{2}}=\frac{2f'(r)}{r}\left(1-\frac{b(r)}{r}\right)-\frac{b(r)}{r^{3}}.
\end{equation}

\subsection{The Shell Region Differential Equations in $\kappa(\mathcal{R},\mathcal{T})-$ gravity}

To find the general equations for the shell region in $\kappa(\mathcal{R},\mathcal{T})$ gravity we can proceed in the following way:\\
using \eqref{EFE1}, we eliminate $\kappa(\mathcal{R},\mathcal{T})$, i.e. we have for $\rho(r)\neq 0$, can always write
\begin{equation}
\kappa(\mathcal{R},\mathcal{T})=\frac{1}{\rho(r)}\frac{b^{\prime}(r)}{r^{2}}.
\end{equation}
Substituting this into Eq. \eqref{nonconservation4}, we obtain the result
\begin{eqnarray}
\frac{d}{dr}\left[\frac{b^{\prime}}{r^{2}}\right]-\frac{b^{\prime}}{r^{2}}\frac{\rho^{\prime}}{\rho}+\frac{b^{\prime}}{r^{2}}\cdot 2f^{\prime}+\frac{b^{\prime}}{r^{2}}\frac{\rho^{\prime}}{\rho(r)} &=& 0\\
\mbox{Or} ~~~~\frac{d}{dr}\Big(\frac{b^{\prime}}{r^{2}}\Big)+ \frac{b^{\prime}}{r^{2}} \cdot 2f^{\prime}(r) &=& 0.
\end{eqnarray}
This differential equation can be integrated to provide
\begin{equation}\label{equationforf(r)}
f(r)=-\frac{1}{2}\ln \bigg( \frac{b^{\prime}(r)}{r^{2}}\bigg).
\end{equation}
This general relation between the metric functions $b(r)$, $f(r)$ should be satisfied for any $\kappa(\mathcal{R},\mathcal{T})$. Inserting now this outcome into Eq. \eqref{combined}, we obtain a differential equation for $b(r)$ as
\begin{small}
\begin{equation}\label{equationforb(r)}
\left[\frac{b^{\prime}(r)}{r^{2}}\right]^{2}=-\frac{1}{r}\frac{d}{dr}\left(\frac{b^{\prime}(r)}{r^{2}}\right)\left(1-\frac{b(r)}{r}\right)-\frac{b(r)b^{\prime}(r)}{r^{5}},
\end{equation}
\end{small}
a non-linear second order differential equation very difficult to solve. To go further, we consider the so-called thin shell approximation.

\subsection{The Thin Shell Approximation}
For this EoS, it is not an easy task to obtain exact analytic solutions. One successful strategy usually employed in the literature is the so-called thin shell approximation, $0<1-b(r)/r\equiv h << 1$. This approximation allow us to set $h\approx 0$ so that one can consider only the lower order terms and we get from \eqref{equationforb(r)}
\begin{equation}
\frac{b^{\prime}(r)}{r^{2}}\left(\frac{b^{\prime}(r)}{r^{2}}+\frac{b(r)}{r^{3}}\right)=0.
\end{equation}
The non-trivial solution for $b(r)$ arises by imposing that 
\begin{equation}
\frac{b^{\prime}(r)}{r^{2}}+\frac{b(r)}{r^{3}}=0.
\end{equation}
The last equation can also be obtained more directly by applying the thin shell approximation to Eq. \eqref{combined}.
Solving for $b(r)$, and using \eqref{equationforf(r)}, we obtain that the metric potentials are given by
\begin{eqnarray}\label{metricpotentials}
b(r)=\frac{D}{r},~~~f(r)=\ln(C_{0}~r^{2}),
\end{eqnarray}
where $D$, $C_{0}$ are constants.

\begin{figure}
\centering
\includegraphics[width=8.3cm,height=5cm]{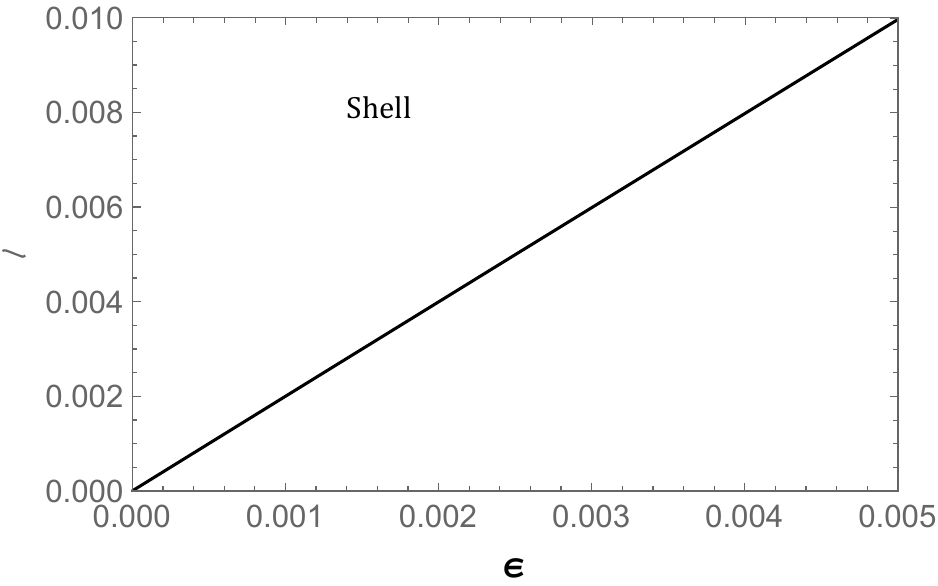}
\caption{Variation of proper length as a function of the thickness within the thin shell.} \label{fig3}
\end{figure}

In the thin shell approximation, the curvature scalar is given by
\begin{equation}
\mathcal{R}=\frac{2b^{\prime}(r)}{r^{2}}+\frac{d}{dr}\left(\frac{b(r)}{r}\right)f^{\prime}(r)=-\frac{6D}{r^{4}}.
\end{equation}
Regarding the proper length of the thin shell,
$e^\lambda$ assumes the following form as
\begin{equation}
e^{\lambda(r)} = \left(1-\frac{b(r)}{r}\right)^{-1}=\left(1-\frac{D}{r^{2}}\right)^{-1}.
\end{equation}
we get the value of the proper length
\begin{eqnarray}
l &=& \int_R^{R+\epsilon} dr \,e^{\lambda/2}=\int_R^{R+\epsilon} {dr \over \sqrt{1-D/r^2}} \nonumber \\
&=& \int_R^{R+\epsilon} {\sqrt{\frac{r^2}{r^{2}-D}}}\,dr=\sqrt{r^{2}-D}\bigg|_R^{R+\epsilon}  \nonumber \\
&=& \sqrt{(R+\epsilon)^{2}-D}-\sqrt{R^{2}-D}.
\end{eqnarray}
This can be rewritten in the form
\begin{equation}
l=(R+\epsilon)\sqrt{1-\frac{D}{(R+\epsilon)^{2}}}-R\sqrt{1-\frac{D}{R^{2}}}.
\end{equation}
If the constant $D$ is small compared to the radius $R$, i.e, $D<<R$ we can approximate the proper length as
\begin{equation}
l\approx \epsilon+\frac{D}{2}\Big(\frac{1}{R}-\frac{1}{R+\epsilon}\Big),
\end{equation}
this further can approximated if $\epsilon<<R$, we finally obtain
\begin{equation}
l\approx \epsilon\Big(1+\frac{D}{2R^{2}}\Big).
\end{equation}
This result means that the proper length of the thin shell is proportional to the thickness $\epsilon$ of the shell. The variation of the proper length as a function of the thickness can be seen in Fig. \ref{fig3}. Another important conclusion is that the proper length of the thin shell does not depend on the particular choice of the $\kappa(\mathcal{R},\mathcal{T})$ running gravitational constant. Now, we discuss the solutions for some specific models.

\subsection{Study of the Thin Shell Solutions for Some Specific Models}
\subsubsection{Model $\kappa(\mathcal{R},\mathcal{T})=8\pi+\beta \mathcal{R}-\alpha \mathcal{T}$}
The density and pressure in the shell for this particular model can be found analytically. Indeed, substituting $\kappa(\mathcal{R},\mathcal{T})=8\pi+\beta \mathcal{R}-\alpha \mathcal{T}$ in \eqref{EFE1}, we find after straightforward manipulations the result
\begin{equation}
p(r)=\rho(r)=\frac{1}{4a} \left[\frac{6bD}{r^{4}}-1+\sqrt{\Big(1-\frac{6bD}{r^{4}}\Big)^{2}-\frac{8ad}{r^{4}}}\right],
\end{equation}
where we have defined the constants $a=\alpha/8\pi$, $b=\beta/8\pi$ and $d=D/8\pi$. In order to represent a feasible physical gravastar, namely, to get positive density in the shell, we have to severely constrain the parameters of the model, i.e, the condition $(1-6bD/r^{4})^{2}>aD/\pi r^{4}$ has to be satisfied, possibly together with $6bD/r^{4}>1$, We consider that with these constraints, the model will become a restrictive an unrealistic one. Therefore, we choose not to pursue a further study of this particular choice, and proceed to discuss now other possible forms of $\kappa(\mathcal{R},\mathcal{T})$.

%%%%%%%%%%%%%%%%%%%%%%%%
%%%%%%%%%%%%%%%%%%%%%%

\subsubsection{Model $\kappa(\mathcal{T})=8\pi-\alpha \mathcal{T}$}
Setting $\beta=0$ in the former case, we obtain
\begin{equation}
p(r)=\rho(r)=\frac{1}{4a} \left(\sqrt{1-\frac{8ad}{r^{4}}}-1\right).
\end{equation}
This solution seems problematic, i.e., since $a$ is a positive constant, $\rho$ seems to be only positive for negative values of $d$, namely, negative values of the constant $D$, and this is not consistent with the thin shell approximation. Indeed, notice that substituting the result $b(r)=D/r$ in the approximation $1-b(r)/r << 1$, it is obtained that $1\approx D/r^{2}$. Hence, if we denote $R$ as the thin shell radius, the last relation implies that $R$ is of order $R\approx \sqrt{D}$, being $D$ necessarily a positive constant. Thus, we conclude that the $\kappa(\mathcal{T})$ model cannot support a gravastar, at least as long as such approximation is valid.

\subsubsection{Model $\kappa(\mathcal{R})=8\pi+\beta \mathcal{R}$}
\begin{figure*}
\centering
\includegraphics[width=8.3cm,height=5cm]{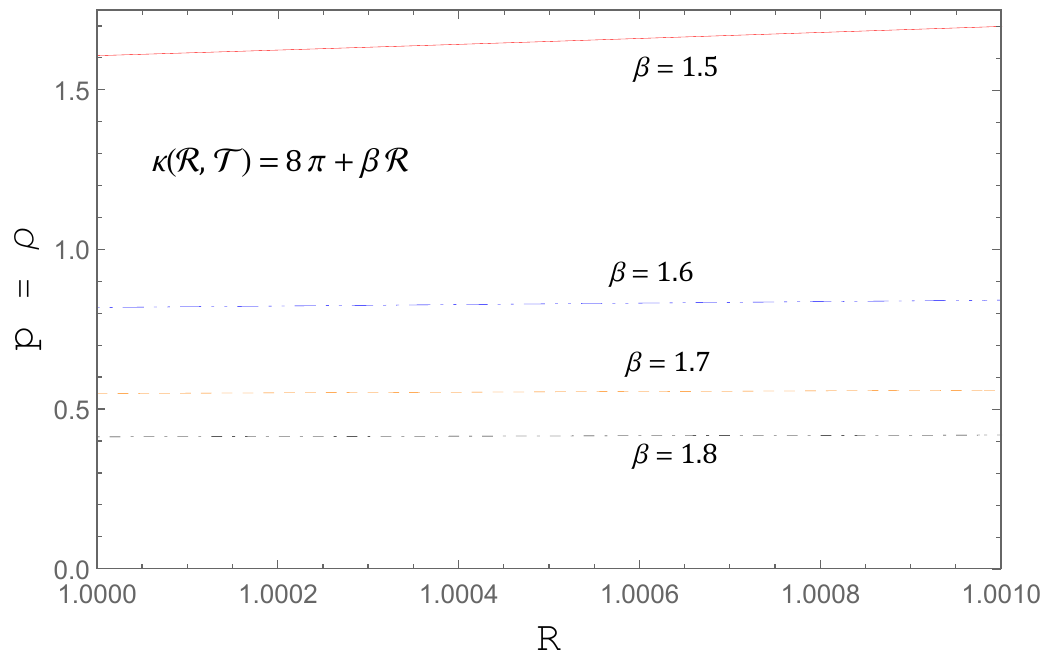}~~\includegraphics[width=8.3cm,height=5cm]{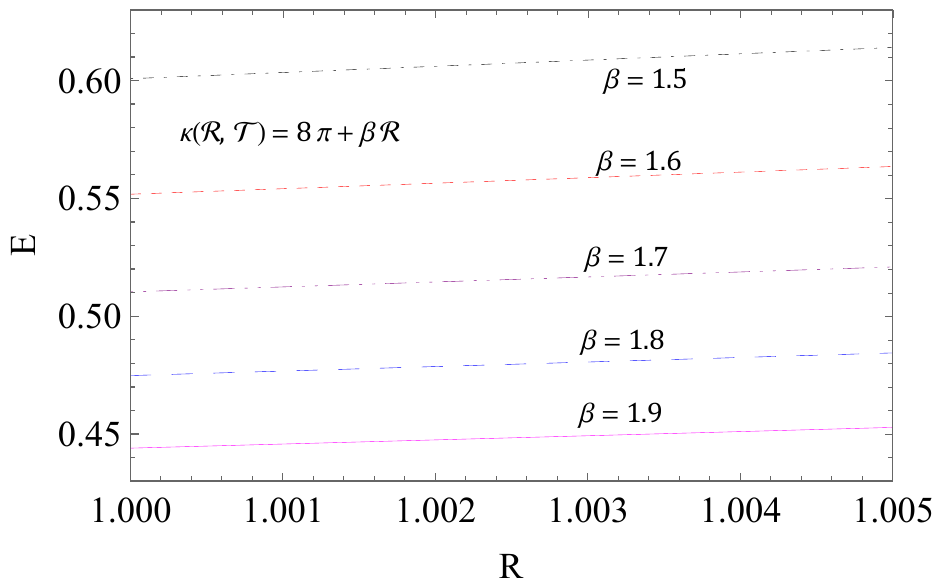}
\caption{Variations of $p=\rho$ and energy within the thin shell for $\kappa(\mathcal{R},\mathcal{T})=8\pi+\beta \mathcal{R}$ by choosing $D=2$.} \label{fig1}
\end{figure*}

The matter density in the model is given by
\begin{equation}
p(r)=\rho(r)=\frac{d}{6bD-r^{4}}. 
\end{equation}
The trend of the $p=\rho$ with respect to $r$ is shown in Fig. \ref{fig1} (left). To be a feasible physical model, i.e, to have positive density, we have to impose the constraint
\begin{equation}
 6bD-r^{4}>0.
 \end{equation}
Then, if we denote $R$ as the radius of the thin shell, we have the following upper bound for the thin shell radius
\begin{equation}
R<\sqrt[4]{6bD}.
\end{equation}
The energy  within the shell for this model is therefore given by 
\begin{eqnarray}
E(r) &=&   \int_R^{R+ \epsilon} 4\pi \rho \,r^2 \,dr= \frac{2\pi  d}{\sqrt[4]{6bD}} \Bigg[\tanh^{-1}\left(\frac{r}{\sqrt[4]{6bD}}\right) \nonumber \\ 
&& \hspace{2cm} -\tan ^{-1}\left(\frac{r}{\sqrt[4]{6bD}}\right)\Bigg]_{R}^{R+\epsilon}.
\end{eqnarray}
The variation of energy in shown in Fig. \ref{fig1} (right).

%%%%%%%%%%%%%%%%%%%%%%%%%%

\begin{figure*}
\centering
\includegraphics[width=8.3cm,height=5cm]{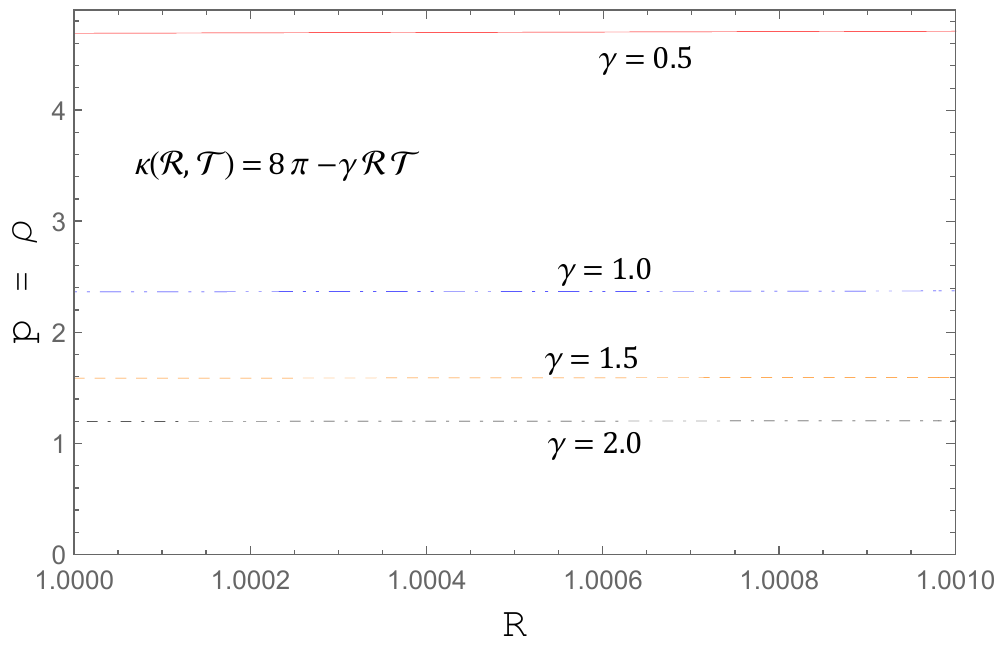}~~\includegraphics[width=8.3cm,height=5cm]{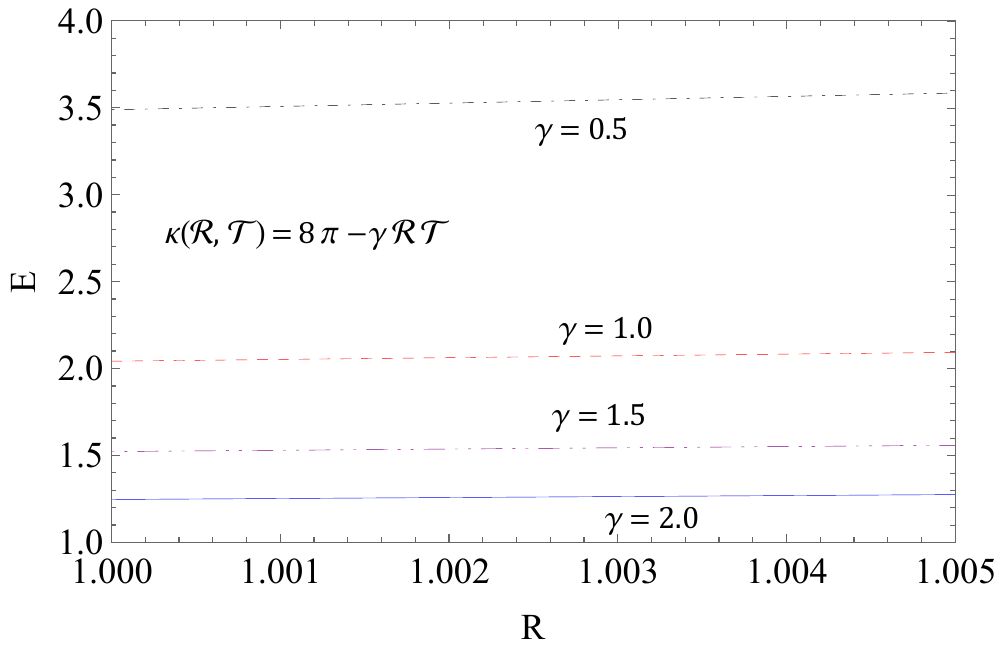}
\caption{Variations of $p=\rho$ and energy within the thin shell for $\kappa(\mathcal{R},\mathcal{T})=8\pi-\gamma \mathcal{R} \mathcal{T}$ by choosing $D=0.9$.} \label{fig2}
\end{figure*}
%%%%%%%%%%%%%%%%%%%%%%%

\subsubsection{Model $\kappa(\mathcal{R},\mathcal{T})=8\pi-\gamma \mathcal{R}\mathcal{T}$}
The matter density and pressure for this specific model that directly couples the matter and curvature trace terms are given by the following expression
\begin{equation}
p(r)=\rho(r)=ar^{4}+\frac{1}{2}\sqrt{(2ar^{4})^{2}+b}.
\end{equation}
Where we have defined the constants $a=\pi/3D\gamma$, $b=1/3\gamma$. Thus, the model $\kappa(\mathcal{R},\mathcal{T})=8\pi-\gamma \mathcal{R}\mathcal{T}$, admits positive density without the need to constrain any free parameter. Again, the variation of $p=\rho$ with respect to $r$ in Fig. \ref{fig2} (left).
On the other hand, the energy  within the shell for this model can be computed by means of the integral 
\begin{eqnarray}
E(r) &=&  \int_R^{R+ \epsilon} 4\pi \rho \,r^2 \,dr = \frac{2 \pi }{3 \gamma D} \Bigg[\frac{2 \pi  r^7}{7}+\frac{\sqrt{\gamma }\, D r^3}{\sqrt{3}}\nonumber \\
&& \hspace{1.5cm} _2F_1\left(-\frac{1}{2},\frac{3}{8};\frac{11}{8};-\frac{4 \pi ^2 r^8}{3 D^2 \gamma }\right)\Bigg]_{R}^{R+\epsilon}.
\end{eqnarray}
where ${}_{2}F_{1}$ is the ordinary hypergeometric function. The variations of energy for $\kappa(\mathcal{R},\mathcal{T})=8\pi+\beta \mathcal{R}$ and $\kappa(\mathcal{R},\mathcal{T})=8\pi-\gamma \mathcal{R} \mathcal{T}$ are shown in Fig. \ref{fig2} (right).

\section{ENTROPY WITHIN THE SHELL}\label{sec6}
\begin{figure*}
\centering
\includegraphics[width=8.3cm,height=5cm]{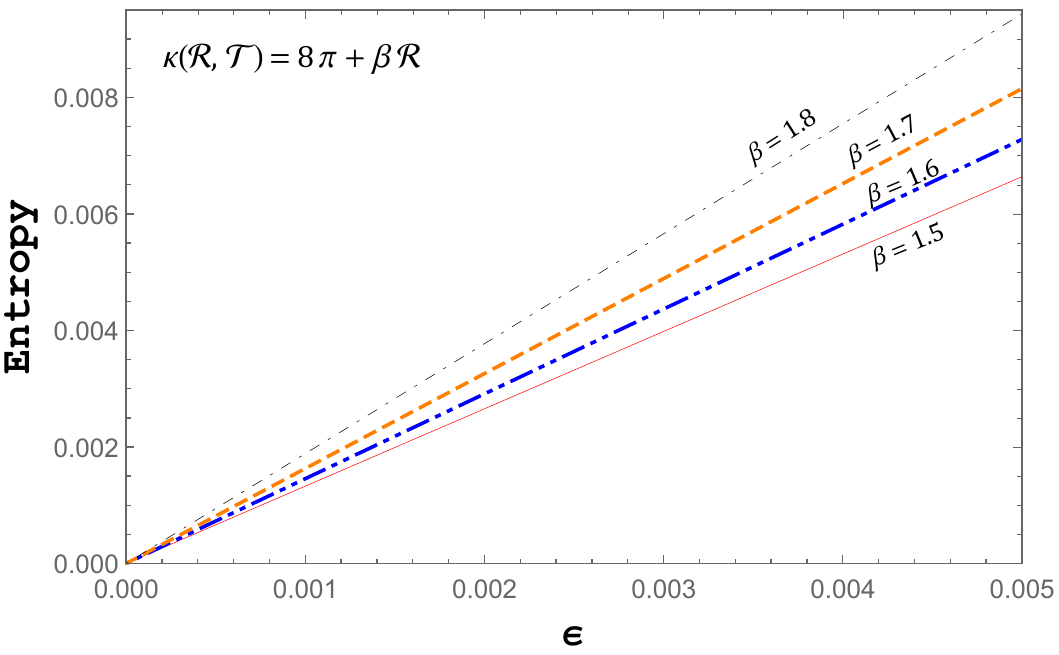}~~\includegraphics[width=8.3cm,height=5cm]{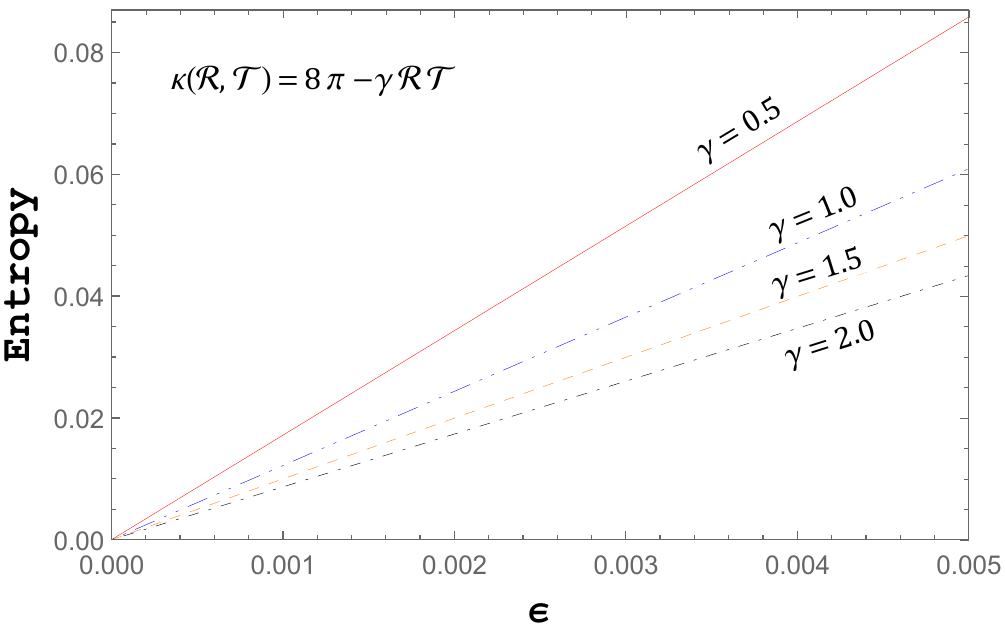}
\caption{Variations of entropy within the thin shell for the two cases ($D=2$ and $D=0.9$ respectively).} \label{fig4}
\end{figure*}

The core region of a gravastar has vanishing entropy density \cite{Mazur01,Mazur04}. However, entropy within the shell is generally not vanishing. For a non-collapsing gravatar, the entropy at the shell is defined as
\begin{equation}
S = \int_R^{R+\epsilon} {4\pi r^2 ~s(r) ~dr\over \sqrt{1-b/r}} ~~~~~\mbox{with}~~~~~s(r) = \chi \sqrt{{p \over 2\pi}}, \label{e60}
\end{equation}
in the unit $\hbar = \kappa_B=1$ and $\chi$, a dimensionless parameter. The thickness of the shell is $\epsilon$. Since \eqref{e60} is non-integrable, we can approximate as follows: consider the primitive integral of \eqref{e60} as
\begin{small}
\begin{equation}
S=\int_R^{R+\epsilon} z'(r) dr =  z(r) \Big|_{R}^{R+\epsilon}=z(R+\epsilon)-z(R) \approx \epsilon \, z'(R)
\end{equation}
\end{small}
at $\epsilon \rightarrow 0$. Hence, \eqref{e60} can be written as
\begin{eqnarray}
S &=& \int_R^{R+\epsilon} {4\pi r^2 ~s(r) ~dr\over \sqrt{1-b/r}} \approx  {4\pi R^2 \, \epsilon~s(R) \over \sqrt{1-b(R)/R}} \nonumber \\
&=& {4\pi R^2 \, \epsilon \, \alpha \over \sqrt{1-b(R)/R}} \, \sqrt{{p(R) \over 2\pi}}~. \label{e62}
\end{eqnarray}
The variations of entropy for $\kappa(\mathcal{R},\mathcal{T})=8\pi+\beta \mathcal{R}$ and $\kappa(\mathcal{R},\mathcal{T})=8\pi-\gamma \mathcal{R} \mathcal{T}$ can be seen in Figs. \ref{fig4}. Entropy in both cases increases linearly with the thickness of the shell $\epsilon$.

%%%%%%%%%%%%%%%%%%%%%%%%%%%%%

\begin{figure*}
\centering
\includegraphics[width=8.3cm,height=5cm]{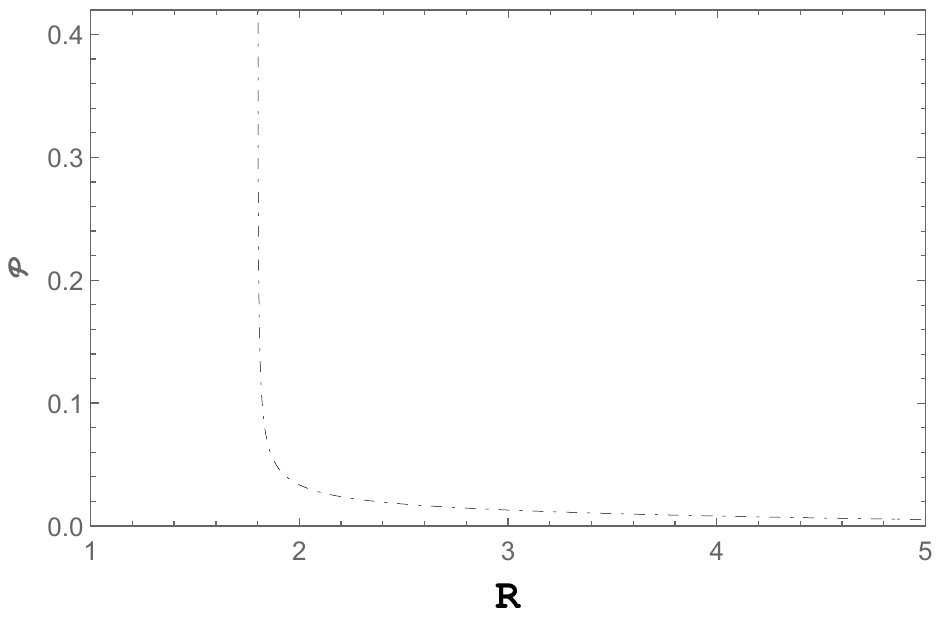}~~\includegraphics[width=8.3cm,height=5cm]{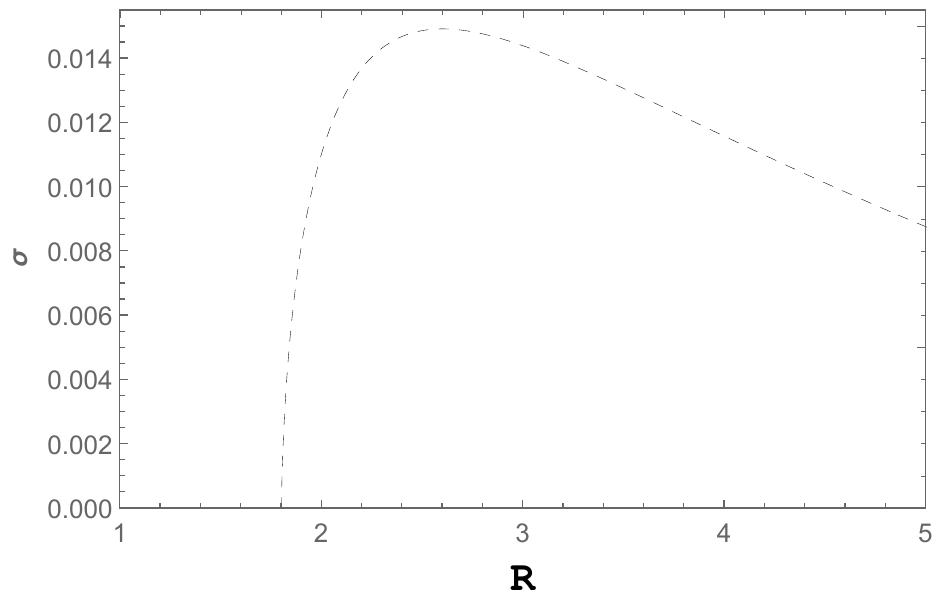}
\caption{Variations of  surface pressure and energy density.} \label{fig5}
\end{figure*}
%%%%%%%%%%%%%%%%%%%%%%%%%

\section{JUNCTION INTERFACE AND SURFACE STRESSES}\label{sec7}
Gravastar has three regions, the interior ($p=-\rho$) filled with dark energy, the shell filled with Zeldovich fluid ($p=\rho$), and an exterior where both pressure and density vanishes ($p=\rho=0$) i.e. a vacuum. The exterior solution is generally accepted as the Schwarzschild solution given by
\begin{equation}
ds^2 = \left[1-{2M \over r}\right]\, dt^2 - {dr^2 \over 1-{2M / r}} -r^2(d\theta^2+\sin \theta \, d\phi^2). \label{eq62}
\end{equation}
At the junctions $r=R$, the interior and the exterior connect smoothly. However, due to the slight mismatch in their derivatives, there arises stress at the junction. The stress tensor is given by the Lanczos equation as
\begin{equation}
\mathcal{S}^i_j = -{1 \over 8\pi} \big(K^i_j-\delta^i_j \, K^q_q \big),
\end{equation}
where $K^{ij}=\mathcal{K}^+_{ij}-\mathcal{K}^-_{ij}$ is the discontinuity in the second fundamental form. Here the signs ``$+$" and ``$-$" correspond to the exterior and the interior regions respectively. The second fundamental at the junction is given by 
\begin{eqnarray}
\mathcal{K}^\pm_{ij}=-n_\nu^\pm \left[{\partial^2 x_\nu \over \partial \xi^i \partial \xi^j}+\Gamma^\nu_{\alpha \beta}\,{\partial x^\alpha \over \partial \xi^i} {\partial x^\beta \over \partial \xi^i} \right]_\Sigma.
\end{eqnarray}

%%%%%%%%%%%%%%%%%%%%%%%%%%%%%%%%
\begin{figure*}
\centering
\includegraphics[width=8.3cm,height=5cm]{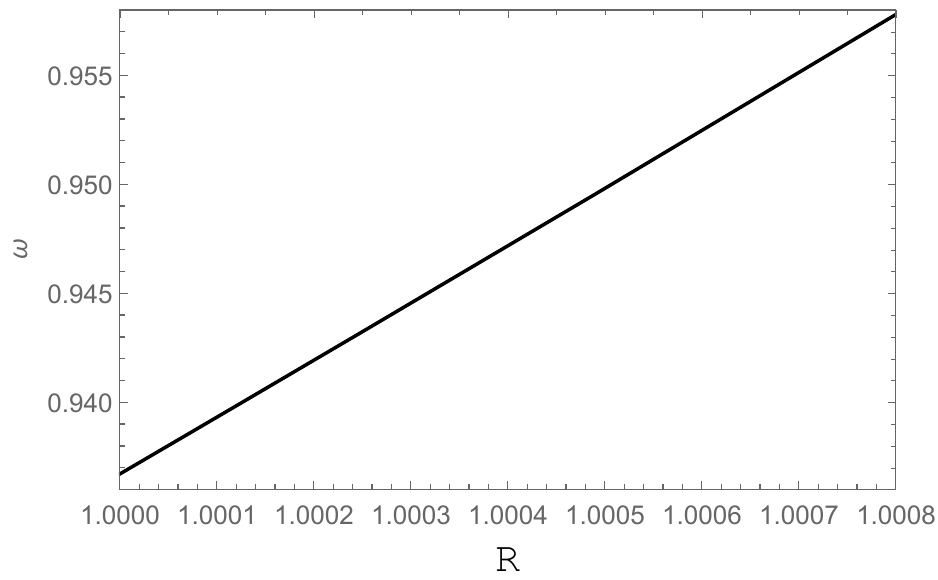}~~\includegraphics[width=8.3cm,height=5cm]{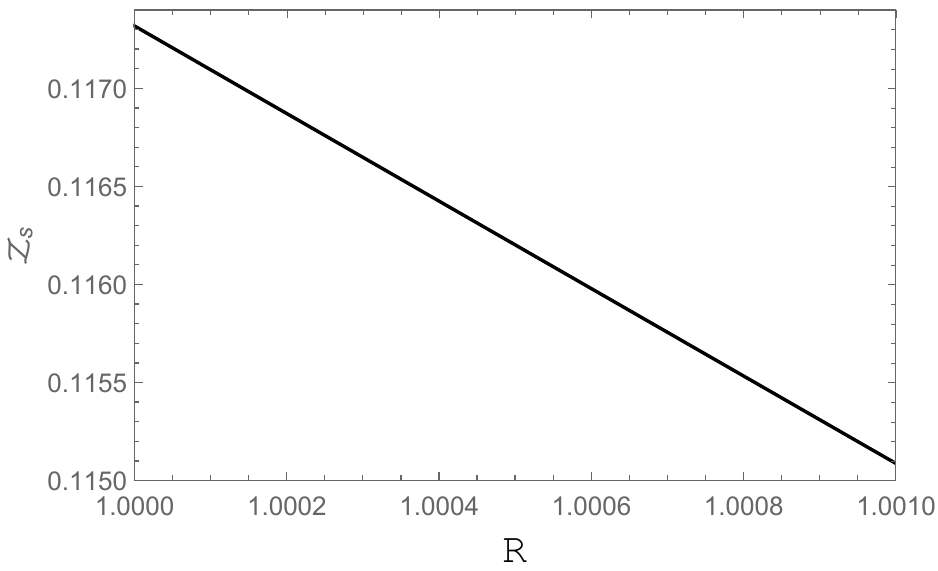}
\caption{Variations of equation of state parameter and  surface redshift $C_0 = 0.895$ and $M = 0.1$.} \label{fig6}
\end{figure*}

\begin{figure}
\centering
\includegraphics[width=8.2
cm,height=5cm]{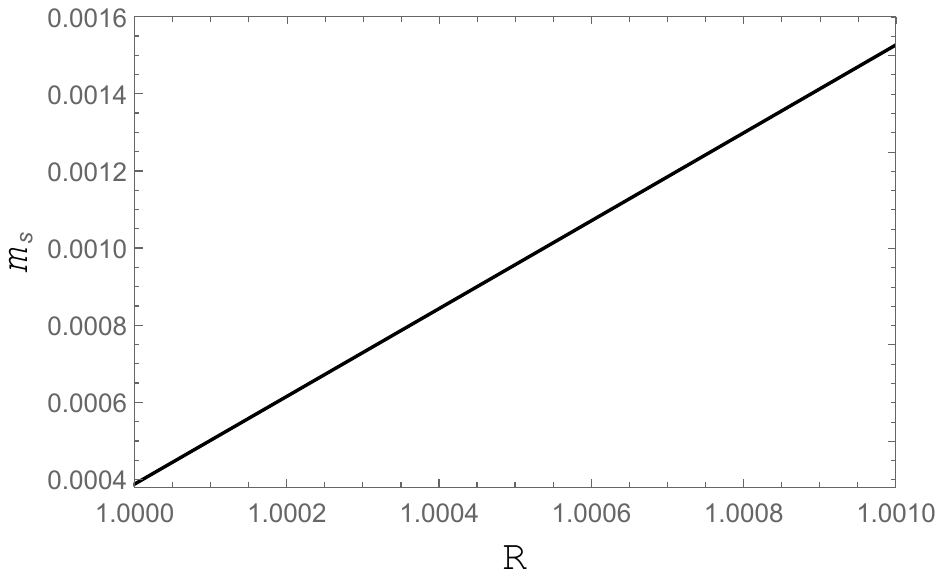}
\caption{Variations of shell mass with respect to shell thickness $C_0 = 0.895$ and $M = 0.1$.} \label{fig7}
\end{figure}
%%%%%%%%%%%%%%%%%%%%%%%%%

Here, $\xi^i$ is the intrinsic coordinates on the shell, and $n_\nu^\pm$ is the unit normals to the surface $\Sigma$. For the exterior spacetime \eqref{eq62}, the unit normal to the surface is given by 
\begin{equation}
 n_\nu^\pm = \pm \left|g^{\alpha \beta} {\partial F \over \partial x^\alpha} {\partial F \over \partial x^\beta} \right|^{-1/2}{\partial F \over \partial x^\nu} ~,~~~~~\forall~~~~n^\mu n_\mu=1.  
\end{equation}
and $F=1-2M/R$. If the surface stress-energy tensor is taken as $\mathcal{S}^i_j=\text{diag}(\sigma,-\mathcal{P},-\mathcal{P},-\mathcal{P})$, where $\sigma$ is the surface energy density and $\mathcal{P}$, the surface pressure, they can be determined as
\begin{eqnarray}
\sigma &=& -{\sqrt{F} \over 4\pi R}\bigg|^+_-  ~~,~~~~~~
\mathcal{P} = -{\sigma \over 2}+{1 \over 16\pi} \left. {F' \over \sqrt{F}}\right|^+_-.
\end{eqnarray}
For the shell, the surface energy density and surface pressure take the form
\begin{eqnarray}
\sigma &=& -{1 \over 4\pi R} \left[\sqrt{1-{2M \over R}}-C_0 R^2 \right],\\
\mathcal{P} &=&  \frac{2-2 M/R-3 C_0 R^2 \sqrt{1-2 M/R}}{16 \pi  R \sqrt{1-2 M/R}}.
\end{eqnarray}
The trends of surface energy and surface pressure are shown in Fig. \ref{fig5}. Now, we can determine the mass of the shell ($m_s$) as
\begin{eqnarray}
m_s = 4\pi R^2 \sigma= R \left[C_0 R^2-\sqrt{1-{2M \over R}} \right],
\end{eqnarray}
from which we can determine the total mass of the gravastar as
\begin{eqnarray}
M = \frac{2 C_0 R^3 m_s-C_0^2 R^6-m_s^2+R^2}{2 R}.
\end{eqnarray}
Further, one can also determine the equation of state parameter at the interface as
\begin{eqnarray}
\omega = {\mathcal{P} \over \sigma}=\frac{3 C_0 R^3 \sqrt{1-2 M/R}+2 M-2 R}{-4 C_0 R^3 \sqrt{1-2 M/R}-8 M+4 R},
\end{eqnarray}
which must have a real value, requires $2M/R<1$ and to avoid singularity $C_0 \neq {\sqrt{1-2M/R}/ R^2}$.\\ 
The variations of the equation of state parameter $\omega$, and shell mass $m_{s}$ can be seen in Fig. \ref{fig6} (left), and Fig. \ref{fig7}.

\section{SURFACE REDSHIFT WITHIN THE THIN SHELL}\label{sec8}
One of the most important physical parameters in gravastar structure analysis is a ratio of wavelengths known as surface redshift $\mathcal{Z}_{s}$. The study of this dimensionless quantity provides valuable information corresponding to the stability and detection of these compact objects. The surface gravitational redshift is defined as $\mathcal{Z}_{s}=\frac{\Delta \lambda}{\lambda_{e}}=\frac{\lambda_{0}}{\lambda_{e}}$, where $\Delta$ represents the fractional change wavelength among the emitted $\lambda_{e}$ and received signal $\lambda_{0}$. For static isotropic matter distribution, Buchdahl \cite{Buch} stated that the value of redshift parameter $\mathcal{Z}_{s}$ should not exceed 2, i.e. $\mathcal{Z}_{s}< 2$. The expression of surface redshift is presented by
\begin{equation}\label{redshift}
\mathcal{Z}_{s}=\vert g_{tt}\rvert ^{-1/2}-1,
\end{equation}
inserting into (\ref{redshift}) the value of the metric potential $g_{tt}$, for the shell region given by Eq. (\ref{metricpotentials}), we get the surface redshift as
\begin{equation}
\mathcal{Z}_{s}=\frac{1}{C_{0}r^{2}}-1 ~~~~\text{with}~~~R \leq r \leq R+\epsilon.
\end{equation}
The variation of surface redshift $\mathcal{Z}_{s}$ as a function of the thickness of the shell is shown in Fig. \ref{fig6} (right).
\section{RESULTS AND DISCUSSION} \label{sec9}
In this current work, we have derived and explored the Einstein field equations describing a gravastar for a generic $\kappa(\mathcal{R},\mathcal{T})$ functional, and then we discussed several special cases corresponding to specific choices of this functional, such as $\kappa(\mathcal{R},\mathcal{T})=8\pi+\beta \mathcal{R}-\alpha \mathcal{T}$, $\kappa(\mathcal{R},\mathcal{T})$ 
$\equiv \kappa(\mathcal{T})=8\pi-\alpha \mathcal{T}$, $\kappa(\mathcal{R},\mathcal{T}) \equiv \kappa(\mathcal{R})=8\pi+\beta \mathcal{R}$ and $\kappa(\mathcal{R},\mathcal{T})=8\pi-\gamma \mathcal{R}\mathcal{T}$. A summary of our findings is the following:
\begin{itemize}
\item In the interior region with EoS $\rho=-p$, the solutions are regular everywhere and, in particular, in the center regardless of the specific form of the $\kappa(\mathcal{R},\mathcal{T})$ functional.
\item For the shell region with EoS $\rho=p$, the specific models, $\kappa(\mathcal{R},\mathcal{T})=8\pi+\beta \mathcal{R}-\alpha \mathcal{T}$, and $\kappa(\mathcal{T})=8\pi-\alpha \mathcal{T}$ provide negative density, therefore they are unphysical solutions and cannot support a gravastar configuration. The viability of the model $\kappa(\mathcal{R})=8\pi+\beta \mathcal{R}$ requires an upper bound for the thin shell radius to get positive density. Nevertheless, the nonlinear model that directly couples the $\mathcal{R}$ and $\mathcal{T}$ traces, i.e. $\kappa(\mathcal{R},\mathcal{T})=8\pi-\gamma \mathcal{R}\mathcal{T}$ supports a gravastar configuration without resorting to any constraint or fine-tuning of the free parameters.
\item Inside the shell, the $p=\rho$ and energy are slightly increasing linearly outward with respect to the shell thickness in the case of $\kappa(\mathcal{R},\mathcal{T})=8\pi+\beta \mathcal{R}$. Both these parameters decrease when the $\kappa(\mathcal{R})-$coupling parameter $\beta$ increases. On the other hand, $p=\rho$ and energy for the case $\kappa(\mathcal{R},\mathcal{T})=8\pi-\gamma \mathcal{R} \mathcal{T}$ are constants throughout the shell and they too decrease when $\kappa(\mathcal{R},\mathcal{T})-$coupling strength $\gamma$ increases.
\item The entropy within the shell increases when the coupling strength $\beta$ increases in $\kappa(\mathcal{R})-$gravity while the same parameter decreases when the coupling strength increases between geometry and matter in nonlinear $\kappa(\mathcal{R},\mathcal{T})-$gravity.
\item The surface redshift $\mathcal{Z}_{s}$ is less than 2, and hence physically inspired.
\end{itemize}

At the end, it can be concluded that the existence of gravastar configurations in $\kappa(\mathcal{R},\mathcal{T})-$gravity strongly depends on the chosen form of the $\kappa-$function. Similar to the parallel competing modified theories of gravity, $\kappa(\mathcal{R},\mathcal{T})$ had already given the wormhole solutions \cite{sark19}, compact star configurations \cite{GRP,Tas23} and now gravastar. Hence, this new theory is becoming a promising new theory of gravity.

\section*{Acknowledgments}

FR, KNS would like to thank the authorities of the Inter-University Centre for Astronomy and Astrophysics, Pune, India for providing research facilities.
 
%\bibliographystyle{plain}
%\bibliography{bib}

\begin{thebibliography}{99}
\bibitem{reig} J. Reignier. The birth of special relativity. \textbf{arXiv:physics/0008229}
\bibitem{Hei} W. Heisenberg,  Zeitschrift für Physik \textbf{33}, 879–893 (1925)
\bibitem{Ein} A. Einstein,  Ann. der Phys. \textbf{17}, 891 (1905)
\bibitem{Hei2} W. Heisenberg,  Zeitschrift für Physik \textbf{43},  172-198 (1927)
\bibitem{hil15}D. Hilbert, Die Grundlagen der Physik.(Erste Mitteilung.). Nachrichten von der Gesellschaft der Wissenschaften zu Göttingen, Mathematisch-Physikalische Klasse \textbf{1915}, 395-408 (1915)
\bibitem{ren07} J. Renn (Ed.). (2007). The genesis of general relativity: Sources and interpretations (Vol. 250). Springer Science and Business Media.
\bibitem{Mazur01}P. Mazur, E. Mottola, \textbf{arXiv:gr-qc/0109035}
\bibitem{Mazur04}P. Mazur, E. Mottola, Proc. Natl. Acad. Sci. USA, \textbf{ 101},  9545 (2004)
\bibitem{visser04} M. Visser, D. L. Wiltshire, Class. Quan. Grav. \textbf{21}, 1135 (2004)
\bibitem{cattoen05} C. Cattoen, T. Faber, M. Visser, Class. Quan. Grav. \textbf{22}, 4189 (2005)

\bibitem{Bilic06} N. Bilic, G. B. Tupper, R. D. Viollier, J. Cosmol. Astropart. Phys. \textbf{02}, 013 (2006)
\bibitem{Lobo07} F. S. N. Lobo, and A. V. B. Arellano, Class. Quan. Grav. \textbf{24}, 1069 (2007)
\bibitem{Rocha08} P. Rocha, R. Chan, M. F A. da Silva, A. Wang, J. Cosmol. Astropart. Phys. \textbf{11}, 010 (2008)
\bibitem{Horvat07} D. Horvat, S. Iliji\'c, Class. Quan. Grav. \textbf{24}, 5637 (2007)

\bibitem{Nandi09} K. K. Nandi, Y. Z. Zhang, R. G. Cai, A. Panchenko, Phys. Rev. D \textbf{79}, 024011 (2009)
\bibitem{Usmani11} A. A. Usmani, F. Rahaman, S. Ray, K. K. Nandi, P. K. F. Kuhfittig, S. A. Rakib, Z. Hasan, Phys. Lett. B \textbf{701}, 388-392 (2011)
\bibitem{FR15} F. Rahaman, S. Chakraborty, S. Ray, A. A. Usmani, S. Islam,  Int. J. Theor. Phys. \textbf{54}, 50-61 (2015) 
\bibitem{Chan}R. Chan, M. F. A. da Silva, P. Rocha, Gen. Rel. Grav. \textbf{43}, 2223–2235 (2011)
\bibitem{C.F.C}C.F.C. Brandt, R. Chan, M.F.A. da Silva, P. Rocha, J. Mod. Phys. \textbf{6}, 879 (2013)
\bibitem{R.CHAN}R. Chan, M.F.A. da Silva, P. Rocha, A. Wang, J. Cosmo. Astropart. Phys. \textbf{03}, 010 (2009)
\bibitem{D.J}D. J. C. Lombardo, C. D. Vigh, Int. J. Mod. Phys. D \textbf{28}, 1950108 (2019)

\bibitem{TH14}T. Harko, F. S.N. Lobo, G. Otalora, E. N. Saridakis, J. Cosmo. Astropart. Phys. \textbf{12}, 021 (2014)
\bibitem{v.c}V. C. de Andrade, L. C. T. Guillen, J. G. Pereira \textbf{arXiv: gr-qc/0011087}
\bibitem{Harko}T. Harko, F. S.N. Lobo, S. Nojiri, S. D. Odintsov, Phys. Rev. D \textbf{84}, 024020 (2011)
 \bibitem{Lovelock}D. Lovelock, J. Math. Phys. \textbf{12}, 498–501 (1971)
\bibitem{N.M.G} N. M. Garc\'ia, F. S N Lobo, J. P. Mimoso, T. Harko, J. Phys.: Conf. Ser. \textbf{314}, 012056 (2011)
\bibitem{MS}M. Sharif, A. Ikram, Eur. Phys. J. C \textbf{76}, 640 (2016)
\bibitem{Y.X}Yixin Xu, T. Harko, S. Shahidi, S.-D. Liang, Eur. Phys. J. C \textbf{80}, 449 (2020)
\bibitem{H.B}H. Barzegar, M. Bigdeli, G. H. Bordbar,  B. E. Panah, Eur. Phys. J. C. \textbf{83}, 151 (2023)
\bibitem{S.G20}S. Ghosh, A. D. Kanfon, A. Das, M. J. S. Houndjo, I. G. Salako, S. Ray, Int. J. Mod. Phys. A \textbf{35}, 2050017 (2020)
\bibitem{S.G21}S. Ghosh, S. Dey, A. Das, A. Chanda, B. C. Paul, J. Cosmo. Astropart. Phys. \textbf{07}, 004 (2021)
\bibitem{A.DAS17}A. Das, S. Ghosh, B. K. Guha, S. Das, F. Rahaman, S. Ray, Phys. Rev. D \textbf{95}, 124011 (2017)
\bibitem{Z.Y20}Z. Yousaf, M.Z. Bhatti, H. Asad, Phys. Dark Uni. \textbf{28}, 100501 (2020)
\bibitem{M.Z20}M. Z. Bhatti, Mod. Phys. Lett. A \textbf{35}, 2050069 (2020)
\bibitem{M.Z.B2020} M. Z. Bhatti, Z. Yousaf, A. Rahaman, Phys. Dark Univ. \textbf{29}, 100561 (2020)
\bibitem{ZY20}Z. Yousaf, Phys. Dark Univ. \textbf{28}, 100509 (2020)
\bibitem{F.S}F. S. N. Lobo, R. Garattini, J. High Ener. Phys. \textbf{12},  065 (2013)
\bibitem{MZ}M. Z. Bhatti, Z. Yousaf, M. Ajmal, Int. J. Mod. Phys. D \textbf{28}, 1950123 (2019)
\bibitem{MZ21}M.Z. Bhatti, Z. Yousaf, T. Ashraf, Chin. J. Phys. \textbf{73}, 167-178 (2021)
\bibitem{ad20}A. Das, S. Ghosh, D. Deb, F. Rahaman, S. Ray, Nucl. Phys. B \textbf{954}, 114986 (2020)
\bibitem{Horvat} D. Horvat, S. Iliji\'c, A. Marunovi\'c, Class. Quan. Grav. \textbf{26}, 025003 (2009)
\bibitem{rahaman}F. Rahaman, A.A. Usmani, S. Ray, S. Islam, Phys. Lett. B \textbf{717}, 1-5 (2012)
\bibitem{ghosh}S. Ghosh, S. Ray, F. Rahaman, B.K. Guha, Ann. Phys \textbf{394}, 230-243 (2018)
\bibitem{Ghosh19}S. Ghosh, D. Shee, S. Ray, F. Rahaman, B.K. Guha, Res. Phys. \textbf{14}, 102473 (2019) 
\bibitem{Sg 19}S. Ghosh , S. Biswas , F. Rahaman , B.K. Guha, S. Ray, Ann. Phys. \textbf{411}, 167968 (2019)
\bibitem{sumita}S. Banerjee, S. Ghosh, N. Paul, F. Rahaman, Eur. Phys. J. Plus, \textbf{135}, 185 (2020)
\bibitem{R.S20}R. Sengupta, S. Ghosh, S. Ray, B. Mishra, S. K. Tripathy, Phys. Rev. D \textbf{102}, 024037 (2020)
\bibitem{S.RAY20}S. Ray, R. Sengupta, H. Nimesh, Int. J. Mod. Phys. D \textbf{29}, 2030004 (2020)
\bibitem{AD16}A. Das, F. Rahaman, B. K. Guha, S. Ray, Eur. Phys. J. C \textbf{76}, 654 (2016)
\bibitem{gr18} G. R. P. Teruel,  Eur. Phys. J. C {\bf 78}, 660 (2018)
\bibitem{ahm22} N. Ahmed, A. Pradhan, Ind. J. Phys. {\bf 96}, 301-307 (2022)
\bibitem{arc22} A. Dixit, A. Pradhan, R. Chaubey,  Int. J. Geom. Math. Mod. Phys. \textbf{19}, 2250013 (2022).
\bibitem{arc23}A. Dixit, S. Gupta, A. Pradhan, A. Beesham,  Symmetry {\bf 15}, 549 (2023) 
\bibitem{sark19} S. Sarkar, N. Sarkar, F. Rahaman, Y. Aditya, To Phys. J. {\bf 2}, 7 (2019)
\bibitem{GRP}G. R. P. Teruel, K. N. Singh, F. Rahaman, T. Chowdhury, Int. J. Mod. Phys. A \textbf{37}, 2250194 (2022)
\bibitem{Tas23} D. Ta\c{s}er, S.S. Do\u{g}ru, Astrophys. Space Sci. \textbf{368}, 49 (2023)
\bibitem{brans1} C.H. Brans, R.H. Dicke, Phys. Rev. {\bf 124(3)}, 925–935 (1961)
\bibitem{brans2}C.H. Brans, \textbf{arXiv:gr-qc/0506063}
\bibitem{Ras}P. Rastall, Phys. Rev. D {\bf 6}, 3357–3359 (1972)
\bibitem{Har}T. Harko, F.S.N. Lobo, S. Nojiri, S.D. Odintsov, \textbf{arXiv:1104.2669}
\bibitem{Lind}L. Lindblom, W.A. Hiscock, J. Phys. A: Math. Gen. {\bf 15}, 1827 (1982)
\bibitem{Viss}M. Visser, Phys. Lett. B \textbf{782}, 83 (2018)
\bibitem{Buch} H.A. Buchdahl, Phys. Rev. {\bf 116}, 1027 (1959)
\bibitem{Hoj}S. A. Hojman, J. Phys. A: Math. Gen \textbf{25}, L291 (1992)
\end{thebibliography}

\end{document}